Cation disorder in stoichiometric MgSnN$_2$ and ambipolar self-doping behavior in off-stoichiometric MgSnN$_2$

Dan Han[*], Stefan S. Rudel, Wolfgang Schnick, and Hubert Ebert

Department of Chemistry, University of Munich, Butenandtstr. 5-13, D-81377 Munich, Germany

Investigations on II-Sn-N$_2$ (II = Mg, Ca) have been started very recently compared to the intense research of the Zn-IV-N$_2$ (IV = Si, Ge, Sn). In this work, we perform a comprehensive study of cation disorder in stoichiometric MgSnN$_2$ and crystal structure characteristic, doping behavior of off-stoichiometric Mg$_{1+x}$Sn$_{1-x}$N$_2$ (x = –0.8, –0.6, –0.5, –0.4, –0.2, 0.2, 0.4, 0.5, 0.6, 0.8) by using the cluster expansion method and first principles calculations. It is found that cation disorder in stoichiometric MgSnN$_2$ induces a band gap reduction because of a violation of the octet rule. Moreover, the local disorder, namely forming (4,0) or (0,4) tetrahedra, would lead to an appreciable band gap reduction and hinder the enhancement of the optical absorption. An off-stoichiometric Mg/Sn ratio can strongly affect the morphology of Mg$_{1+x}$Sn$_{1-x}$N$_2$ samples due to the higher ionicity of the Mg-N bonds in comparison with Zn-N bonds. Furthermore, Mg$_{1+x}$Sn$_{1-x}$N$_2$ compounds show an ambipolar self-doping behavior, i.e., Mg-rich Mg$_{1+x}$Sn$_{1-x}$N$_2$ show p-type doping while Sn-rich ones exhibit n-type doping owing to the formation of acceptor-type antisite defect Mg$_{Sn}$ or donor-type antisite defect Sn$_{Mg}$, respectively.

# I. INTRODUCTION

Binary nitride semiconductors and related alloys have wide applications, such as high-power and light emitting devices [1-5]. As analogues to binary nitride semiconductors, ternary nitride semiconductors have recently drawn immense attention because of their richer composition space and potentially wider applications [6-11]. Among them, the Zn-IV-$N_2$ (IV = Si, Ge, Sn) series and their alloys have been under intense exploration [12-16]. Due to the beneficial characteristics, viz. earth abundant composition, suitable band gap and benign defect properties [6, 9, 12], $ZnSnN_2$ has gained special interest and has been proposed as a potential photovoltaic absorber [9]. However, it was lately identified that $ZnSnN_2$ suffers from a degenerate electron concentration ($10^{18}$–$10^{21}$ cm$^{-3}$) no matter whether synthesized as bulk by a high-pressure metathesis reaction or as thin films grown by molecular beam epitaxy (MBE) or sputtering [8, 9, 17-19], which suggests that $ZnSnN_2$ may be unsuitable for photovoltaic application for which the carrier concentration have to be around $10^{16}$–$10^{18}$ cm$^{-3}$ [20]. Although $ZnSnN_2$ based photovoltaic devices have been fabricated, yet they show low power conversion efficiency of 1.54% [21, 22]. Nevertheless, solutions to reduce the carrier density in $ZnSnN_2$ are still under investigation because of its aforementioned intrinsic advantages [23-25]. Such a degenerate behavior is also found in $ZnGeN_2$ [26, 27] but a corresponding finding has not been reported yet for $ZnSiN_2$.

In contrast to the intense research on the Zn-IV-$N_2$ (IV = Si, Ge, Sn), studies on II-Sn-$N_2$ (II = Mg, Ca) have been started only recently [28-34]. The electronic properties of $MgSnN_2$ have been studied first by Lambrecht *et al.* [28], with a band gap predicted close to 3.4 eV while their recent work revised the band gap of $MgSnN_2$ to be 2.3 eV after including Sn-4d semicore states [29]. Nonetheless, the cation disorder of $MgSnN_2$ remains insufficiently explored, which is expected to have an important effect on the band gaps according to preceding studies of Zn-IV-$N_2$ (IV = Si, Ge, Sn), i.e., cation disorder strongly influences their electronic structure [14, 35]. Experimentally, Greenaway *et al.* have prepared $MgSnN_2$ thin films with a broad range of cation compositions and gave a phase map of samples under different Mg/Sn ratios and temperatures [31]. Apart from the wurtzite-type phase, this map shows in particular that a rocksalt-type phase appears under Mg-rich conditions, which has not been observed in ZnSnN$_2$ and ZnGeN$_2$ under Zn-rich conditions [24, 36]. Additionally, MgSnN$_2$ crystallizes in the rocksalt-type structure by using a metathesis reaction under high pressure whereas ZnSnN$_2$ crystallizes in the wurtzite-type structure when using the same preparation method [32, 33]. This leads to the question why MgSnN$_2$ can appear in the rocksalt-type phase while ZnSnN$_2$ and ZnGeN$_2$ appear only in the wurtzite-type phase. It is still an open question to date, which will be addressed in this contribution.

In this work, we perform a comprehensive study of stoichiometric MgSnN$_2$ and off-stoichiometric Mg$_{1+x}$Sn$_{1-x}$N$_2$ (x = −0.8, −0.6, −0.5, −0.4, −0.2, 0.2, 0.4, 0.5, 0.6, 0.8) by using the cluster expansion method and first principles calculations. We find that cation disorder in stoichiometric samples induces a band gap reduction because of the formation of motifs violating the octet rule, which is similar to the cases of ZnSnN$_2$ and ZnGeN$_2$. Furthermore, partial disorder cannot be avoided but a fully disordered state is difficult to be obtained for MgSnN$_2$ under nonequilibrium growth conditions. An off-stoichiometric Mg/Sn ratio can strongly affect the morphology of Mg$_{1+x}$Sn$_{1-x}$N$_2$ due to the high ionicity of Mg-N bonds. In addition, Mg$_{1+x}$Sn$_{1-x}$N$_2$ shows an ambipolar self-doping behavior, i.e., Mg-rich samples show p-type doping while Sn-rich samples exhibit n-type doping.

## II. CALCULATION METHODOLOGY

The β-NaFeO$_2$ structure type with the space group *Pna*2$_1$ (no. 33) was considered for fully ordered (FO) MgSnN$_2$ in the wurtzite-type phase, and space group *P*2/*c* (no. 13) was used for the rocksalt-type phase, as suggested by Ref. [31]. Both crystal structures obey the local octet rule. Different levels of cation disorder for stoichiometric MgSnN$_2$ were considered (see Table I), which starts from the ideal *Pna*2$_1$ crystal structure because it was identified as the most stable crystal structure [31]. Accordingly, 128-atom supercells with different levels of cation disorder were generated by the cluster expansion method [37, 38] by using the *mcsqs* code implemented in the Alloy Theoretic Automated Toolkit (ATAT) [39]. Two shells for pairs and triplets were considered. For off-stoichiometric $Mg_{1+x}Sn_{1-x}N_2$ compounds, their special quasi-random structures (SQS) with various Mg/Sn concentrations (x = −0.8, −0.6, −0.5, −0.4, −0.2, 0.2, 0.4, 0.5, 0.6, 0.8) were generated by using the same method. A 32-atom supercell was used for $Mg_{1+x}Sn_{1-x}N_2$ (x = −0.5, 0.5), and an 80-atom supercell was employed for the other compositions.

The geometry optimization, calculations of total energy and electronic structure of ordered and disordered structures were performed within the framework of density functional theory via the Vienna Ab initio Simulation Package (VASP) [40-42]. The projected augmented wave pseudopotential was utilized to describe the ion-electron interaction [43, 44]. The following electron configurations were treated as valence electrons: Mg ($2p^63s^2$), Sn ($4d^{10}5s^25p^2$), N ($2s^22p^3$) and Zn ($3d^{10}4s^2$). A 520 eV energy cutoff of a plane-wave basis was used for all calculations. The convergence criterion for the energy and the force were set at $10^{-5}$ eV and 0.01 eV/Å, respectively. The Monkhorst-Pack k-point mesh with a grid spacing of $\sim 2\pi \times 0.03$ Å$^{-1}$ was adapted for the Brillouin zone integration. The Perdew-Burke-Ernzerhof exchange-correlation functional revised for solids (PBEsol) within the generalized gradient approximation (GGA) was employed [45]. GGA typically underestimates the band gap. To correct the underestimation of the band gap, the Becke and Johnson (BJ) potential as modified by Tran and Blaha, generally known as mBJ was employed, which can give comparable results at relatively low expense with costly functionals such as hybrid functional and the quasiparticle self-consistent GW approximation [46]. The parameters *α* and *β* as suggested in Ref. [46] were used. We made a comparison between the band gaps calculated by the mBJ functional and GW approximation, as discussed in Sec. IIIA. The absorption coefficient spectra of fully ordered and two partially disordered stoichiometric MgSnN$_2$ samples were calculated using a denser k-point mesh (less than $2\pi \times 0.01$ Å$^{-1}$).

## III. STOICHIOMETRIC MgSnN$_2$

### A. Electronic structure of fully ordered MgSnN$_2$

It was reported that MgSnN$_2$ crystallizes in the wurtzite phase when grown by the sputtering method [31, 34]. Besides the wurtzite-type crystal structure, MgSnN$_2$ can crystallize in the rocksalt-type phase by a metathesis reaction under high pressure [32, 33] even though the rocksalt-type phase is energetically much less favorable than the wurtzite-type phase [31]. For this reason, we calculated the band structures for both phases, namely, $Pna2_1$ (no. 33) for the wurtzite and $P2/c$ (no. 13) for the rocksalt-type phase. As shown in Fig. 1, MgSnN$_2$ in the $Pna2_1$ phase exhibits a direct band gap (2.29 eV) at the Γ point while MgSnN$_2$ in the $P2/c$ phase shows an indirect band gap (2.82 eV) with valence band maximum (VBM) at a point along the P-N path and conduction band minimum (CBM) at the Γ point. Both phases show a common characteristic, i.e., the dispersion of the conduction band is large whereas the dispersion of the valence band is relatively small, suggesting potentially better electron transport than hole transport. Our mBJ calculated results for the gap agree with previous GW calculations [29, 31] but they are both larger than the corresponding experimental values ( ~2.0 eV for the wurtzite and 2.3 eV for the rocksalt-type phase) [31, 33]. One possible reason for the disparity between the experimental band gaps and calculated ones can be the band gap reduction induced by the cation disorder in the synthesized samples.

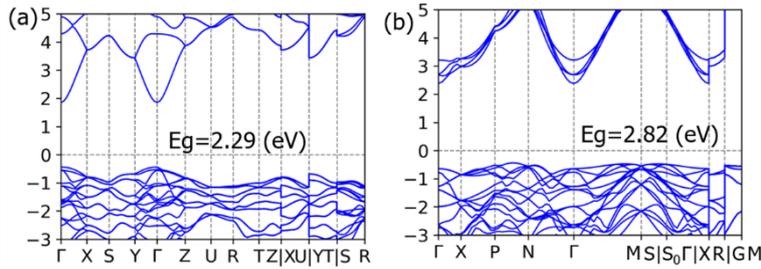

FIG. 1. Band structures of MgSnN$_2$ in the (a) wurtzite phase (space group $Pna2_1$) and rocksalt phase (space group $P2/c$), respectively.

### B. Cation disorder in stoichiometric MgSnN$_2$

The commonly observed decreased band gap of ZnSnN$_2$ and ZnGeN$_2$ thin films is related to cation disorder [35, 47, 48]. For the fully ordered ternary nitrides Zn-IV-N$_2$ (IV = Si, Ge, Sn) in the wurtzite-type phase, the nitrogen atom in each tetrahedron is exactly coordinated with two group-II and two group-IV atoms (hereafter called a (2,2) tetrahedron), which preserve the charge neutrality and obey the octet rule locally. It was reported that cation disorder leads to the formation of some tetrahedra having one group-II and three group-IV atoms (1,3) or three group-II and one group-IV atoms (3,1), which violate the octet rule locally but maintain the global charge neutrality by forming (3,1) and (1,3) tetrahedra elsewhere in the crystal structure [48, 49]. Moreover, (4,0) or (0,4) tetrahedra which only contain four group-II or four group-IV atoms could appear in the disordered samples. We expected that cation disorder has a similar effect on the band gap of $MgSnN_2$ to $ZnSnN_2$ and $ZnGeN_2$, and we studied the impact of different levels of cation disorder in stoichiometric $MgSnN_2$. These are quantified by the ratio of tetrahedra violating the octet rule to the total number of tetrahedra (global disorder), i.e., cation disorder on the level of 17.19%, 12.50%, 18.75%, 34.38%, 43.75%, 57.82% corresponds to six disordered configurations. For simplicity, we refer six disordered configurations with aforementioned levels of cation disorder to P-1, P-1-swap, P-2, P-3, P-4 and fully disorder (FD). Note that the P-1-swap configuration was derived from the P-1 configuration but its (4,0) tetrahedron was eliminated manually.

As shown in Table I, configurations with an increased degree of cation disorder exhibit larger relative energies, that is, the energy difference between the P-1-swap configuration and the FO configuration shows the smallest value (14 meV/atom), while that between the FD configuration and the FO configuration is the largest (126 meV/atom). Partially disordered configurations without (4,0) or (0,4) tetrahedra are likely to appear under nonequilibrium growth conditions because of the small energy difference. In contrast, the large energy difference between the FD and the FO configuration indicates that the FD sample is difficult to form since it is energetically too costly. Overall, it would be challenging to synthesize $MgSnN_2$ samples with full long-range ordering and difficult to obtain fully disordered samples as well.

TABLE I. Numbers of tetrahedra of different motifs in fully ordered (FD), partial disordered and fully disordered (FD) configurations of stoichiometric $MgSnN_2$ and energy with respect to $MgSnN_2$ in the FO wurtzite-type phase ($\Delta E$).

| Configuration | (2,2) | (3,1) | (1,3) | (4,0) | (0,4) | (4,1) | $\Delta E$ (meV/atom) |
|---|---|---|---|---|---|---|---|
| P-1 | 53 | 4 | 6 | 1 | 0 | 0 | 25 |
| P-1-swap | 56 | 4 | 4 | 0 | 0 | 0 | 14 |
| P-2 | 52 | 5 | 7 | 1 | 0 | 0 | 33 |
| P-3 | 42 | 11 | 11 | 0 | 0 | 0 | 37 |
| P-4 | 36 | 16 | 9 | 1 | 1 | 1 | 85 |
| FD | 27 | 15 | 12 | 4 | 6 | 0 | 126 |
| FO | 64 | 0 | 0 | 0 | 0 | 0 | 0 |

Besides the relative energies of the disordered configurations, the impact of varying the degree of cation disorder on the band gap is further investigated. Figure 2 shows the trend of the band gap as the degree of disorder varies. The band gap decreases as the degree of cation disorder increases, as can been seen from the FO, P-1, P-2, P-4 and FD configurations, for which the global disorder is increased gradually (See the middle panel of Fig. 2). Among the various configurations, the FD configuration with the highest level of global disorder shows a complete closing of the band gap. Notably, besides the global disorder, local disorder (appearance of (4,0) or (0,4) tetrahedra) seems to play a role on the band gap reduction. The P-2 configuration shows a smaller band gap than the P-3 configuration even though the former has a lower level of global disorder than the latter, as shown in the top and middle panel of Fig. 2. We note that there is a (4,0) tetrahedron in the P-2 configuration while no (4,0) or (0,4) tetrahedra in P-3 configuration (see Table I), which means that P-2 configuration exhibits higher level of local disorder than P-3 configuration. Moreover, if the (4,0) or (0,4) tetrahedra were eliminated, the band gap would increase, as it is observed from the comparison between the band gaps for the P-1 and the P-1-swap configuration in Fig. 2.

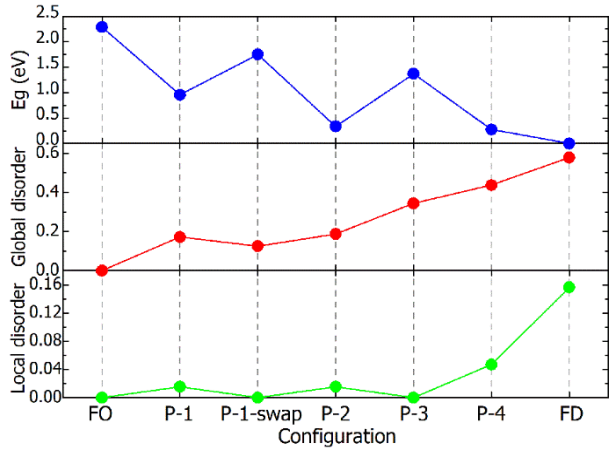

FIG. 2 Band gaps, degree of global and local disorder of fully ordered (FO), partially disordered and fully disordered (FD) configurations of stoichiometric MgSnN$_2$.

To figure out how the local disorder influences the band gap of disordered MgSnN$_2$ in detail, we analyze the total density of states of configurations without (4,0) and (0,4) tetrahedra and with (4,0) or (0,4) tetrahedra, respectively. Compared to the FO configuration, as shown in Fig. 3, cation disorder predominantly affects the valence band edge (VBE) of the P-1-swap and the P-3 configurations containing no (4,0) or (0,4) tetrahedra, i.e., a defect state with low density appears around the VBE. For the P-1 and P-2 configurations containing a (4,0) tetrahedron, there is an isolated defect state, as marked by an arrow in Fig. 3b. This local defect state leads to a large decrease of the band gap for the P-1 and P-2 configurations, viz. 1.33 eV and 1.95 eV, respectively. In FD configuration, there exists a continuum of defect states throughout the original band-gap region, and thus entailing the band gap closed. A similar behavior is observed for the fully disordered ZnGeN$_2$ [49].

Figure 4a shows the charge density contours for the VBM and the CBM of the FO configuration, of which the VBM is mainly contributed by N-2p orbitals and the CBM is composed of N-2s and Sn-5s antibonding states as revealed by previous studies as well [28, 50]. The charge density related to the VBM of P-1-swap configuration localizes at the N atom centering in the (3,1) tetrahedron and its first-neighbor (2,2) tetrahedra, whereas the charge density of the VBM of the P-1 configuration mainly localizes at the N atom centering in the (4,0) tetrahedron, as displayed in Fig. 4b and Fig. 4c, respectively. The localization of charge density for VBM of P-1 configuration reflects the isolated local defect state aforementioned. Unlike the charge density distribution of the VBM, the charge density distribution of the CBM of P-1-swap and P-1 configurations extends the whole structure, indicating that the conduction bands are less perturbed.

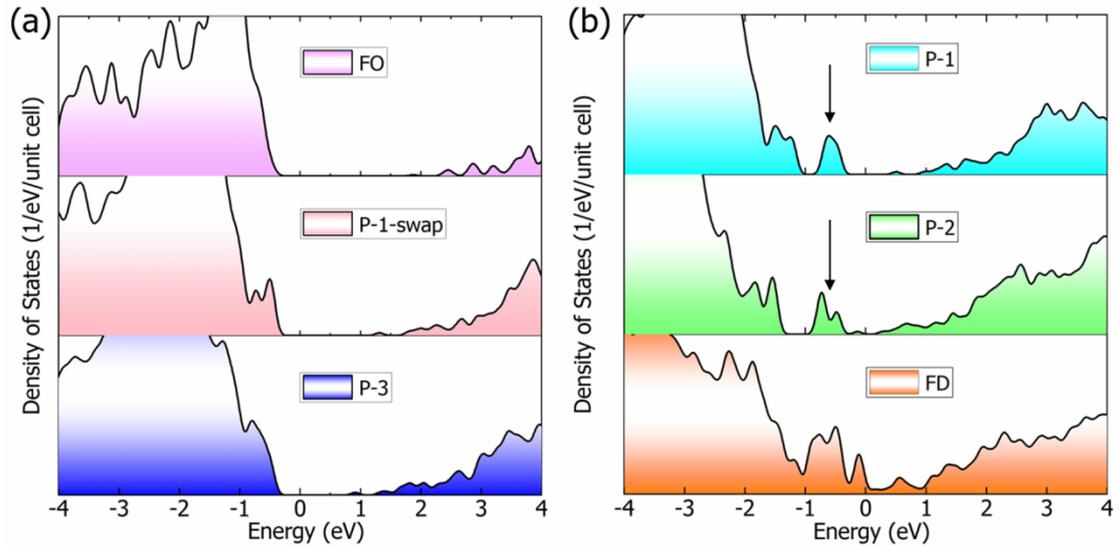

FIG. 3. Total density of states of (a) FO configuration and two cation-disordered MgSnN$_2$ configurations without (4,0) or (0,4) motif (P-1-swap and P-3) and (b) three cation-disordered MgSnN$_2$ configurations with (4,0) motif (P-1, P-2 and FD).

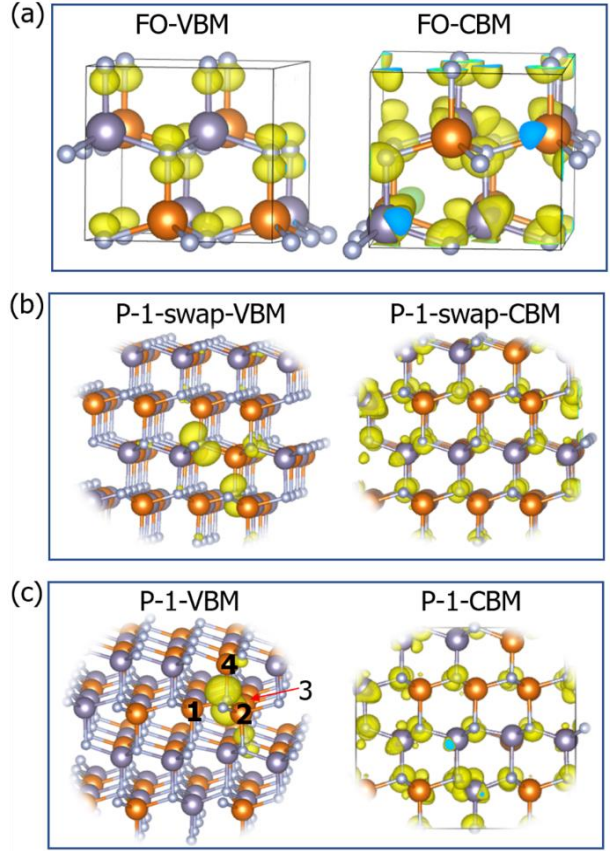

FIG. 4. Charge density contours of VBM and CBM for (a) FO, (b) P-1-swap and (c) P-1 configuration. The isosurface values are 0.01 e/Å$^3$ for the VBM and 0.005 e/Å$^3$ for the CBM of the FO configurations, and 0.005 e/Å$^3$ for the VBM and 0.0005 e/Å$^3$ for the CBM of the P-1-swap and P-1 configurations.

Ternary nitrides such as ZnSnN$_2$ have been suggested as photovoltaic absorbers which require a high absorption coefficient within the visible light regime. Considering the inevitable partial cation disorder in MgSnN$_2$, we further studied the effect of cation disorder on the absorption spectrum. As shown in Fig. 5, the absorption coefficient of the FO configuration reaches $10^4$ cm$^{-1}$ within the 0.33 eV energy range to the onset, and then it reaches $10^5$ cm$^{-1}$ within the next 0.98 eV. For the P-1 and P-1-swap configurations, the absorption onsets shift to a lower energy because of the decreased band gaps. The behavior of optical absorption for the P-1-swap configuration is similar to that for the FO configuration. However, the P-1 configuration shows a rather different behavior for the optical absorption, i.e., its absorption coefficient reaches $10^4$

cm$^{-1}$ within a large energy range (0.82 eV) to the absorption onset and maintains almost the same order of magnitude within the following energy range of 0.50 eV, and then it reaches 10$^5$ cm$^{-1}$ slowly. The slow increase of the absorption coefficient should be ascribed to the isolated defect states in the P-1 configuration. Altogether, we can see that cation disorder results in a narrowing of the band gap and shifting the absorption onset to a lower energy, which widens the absorption energy range in visible light spectrum. This behavior is expected to be beneficial for photovoltaic applications. Nevertheless, it is worth noting that the isolated local defect states influence the enhancement of the optical absorption because they perturb the band edge strongly, and thus the formation of (4,0) and (0,4) tetrahedra should be avoided during the preparation as far as possible.

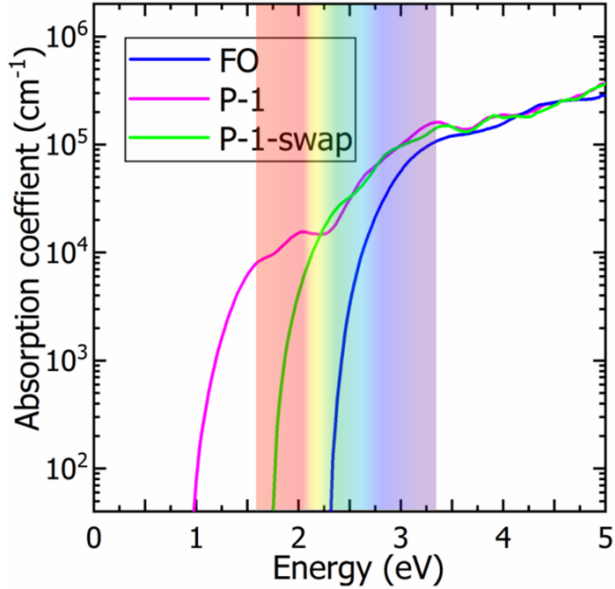

FIG. 5. Calculated absorption spectra of the FO, P-1 and P-1-swap configurations.

### IV. OFF-STOICHIOMETRIC Mg$_{1+x}$Sn$_{1-x}$N$_2$

#### A. Crystal structure characteristic

As displayed in the phase map of Mg$_{1+x}$Sn$_{1-x}$N$_2$ (−0.8 ≤ x ≤ 0.8) given by Greenaway *et al.* experimentally [31], Mg$_{1+x}$Sn$_{1-x}$N$_2$ samples show poor crystallinity under extreme Mg-poor (x < −0.6) or extreme Mg-rich (x > 0.7) conditions in combination with low temperature; samples in the Mg-rich region (0.5 < x < 0.7) show the occurrence of two phases (wurtzite and rocksalt-type) under low temperature; while for the remaining region of concentration and under low temperature, $Mg_{1+x}Sn_{1-x}N_2$ samples are in the wurtzite phase. Theoretically, we consider the composition x = −0.8, −0.6, −0.5, −0.4, −0.2, 0.2, 0.4, 0.5, 0.6, 0.8 for off-stoichiometric $Mg_{1+x}Sn_{1-x}N_2$ compounds. Based on the crystal structures after geometry optimization, we find that the Mg/Sn ratio indeed largely affects the crystal structures, i.e., $Mg_{1+x}Sn_{1-x}N_2$ with x = −0.8, −0.6, −0.5 and 0.8 show poor crystallinity; the ones with x = −0.4, −0.2, 0.2 exhibit wurtzite-like crystal structures; those with x = 0.4, 0.5, 0.6 have mixed crystal structures of which each N atom coordinates with five cationic atoms and each cationic atom bonds to five N atoms. Our theoretical discussion of crystal structure qualitatively agrees with the experimental observation of Greenaway *et al.* despite slight deviations of the Mg/Sn ratio boundary from the experimental values.

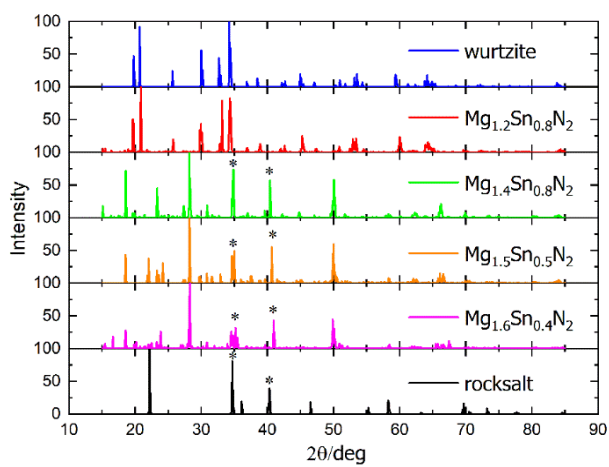

FIG. 6. Simulated powder XRD patterns (Cu Kα) of $MgSnN_2$ in the wurtzite and rocksalt-type phase, and for $Mg_{1+x}Sn_{1-x}N_2$ with x = 0.2, 0.4, 0.5, 0.6.

Considering that the Mg-rich $Mg_{1+x}Sn_{1-x}N_2$ compounds with x = 0.4, 0.5, 0.6 show non-tetra-coordination, we calculated their X-ray diffraction (XRD) patterns (Cu Kα). The XRD patterns of $MgSnN_2$ in the wurtzite and rocksalt-type phase, and $Mg_{1.2}Sn_{0.8}N_2$ are displayed in Fig. 6 as well. As can be seen, $Mg_{1.2}Sn_{0.8}N_2$ exhibits a similar XRD pattern as $MgSnN_2$ in the wurtzite-type phase except for the slightly different intensity of the peaks at 32.75° and 34.35°. $Mg_{1.4}Sn_{0.6}N_2$, $Mg_{1.5}Sn_{0.5}N_2$ and $Mg_{1.6}Sn_{0.4}N_2$ show two characteristic diffraction peaks similar to rocksalt-type $MgSnN_2$, which are marked by stars in Fig. 6.

As mentioned above, $ZnSnN_2$ does not crystallize in the rocksalt-type structure no matter under which kind of synthesis conditions [32]. Moreover, there is no rocksalt-type phase appearing in the phase map of the Zn-rich region [24]. We theoretically investigated the crystal structures of $Zn_{1.4}Sn_{0.6}N_2$, $Zn_{1.5}Sn_{0.5}N_2$ and $Zn_{1.6}Sn_{0.4}N_2$. In contrast to the $Mg_{1+x}Sn_{1-x}N_2$ counterparts, $Zn_{1.4}Sn_{0.6}N_2$, $Zn_{1.5}Sn_{0.5}N_2$ and $Zn_{1.6}Sn_{0.4}N_2$ show the wurtzite-like structure, as depicted in Fig. 7. This different behavior can be ascribed to the ionicity of the bond, i.e., the stronger the ionic bond is, the more likely it is for the compound to form a structure with a high degree of coordination [32]. Here, electron localization function (ELF) is employed to describe the bonding characteristic. As shown in Fig. 8a and b, electrons are more delocalized around the Mg atom than the Zn atom. The 1D profile along the Mg-N bond (Fig. 8c) shows the characteristic of mixed ionic-covalent bonding [51, 52], i.e., the ELF minimum is above zero in the non-nuclear region (> 0.05). The 1D profile along the Zn-N bond (Fig. 8d) shows one basin between the core of Zn and core of N, and ELF maximum moved to the position of the N atom because of larger electronegativity of N, which shows the characteristic of covalent bonding. Accordingly, $ZnSnN_2$ forms a tetra-coordinated structure while $MgSnN_2$ is able to form highly coordinated structures especially under Mg-rich conditions.

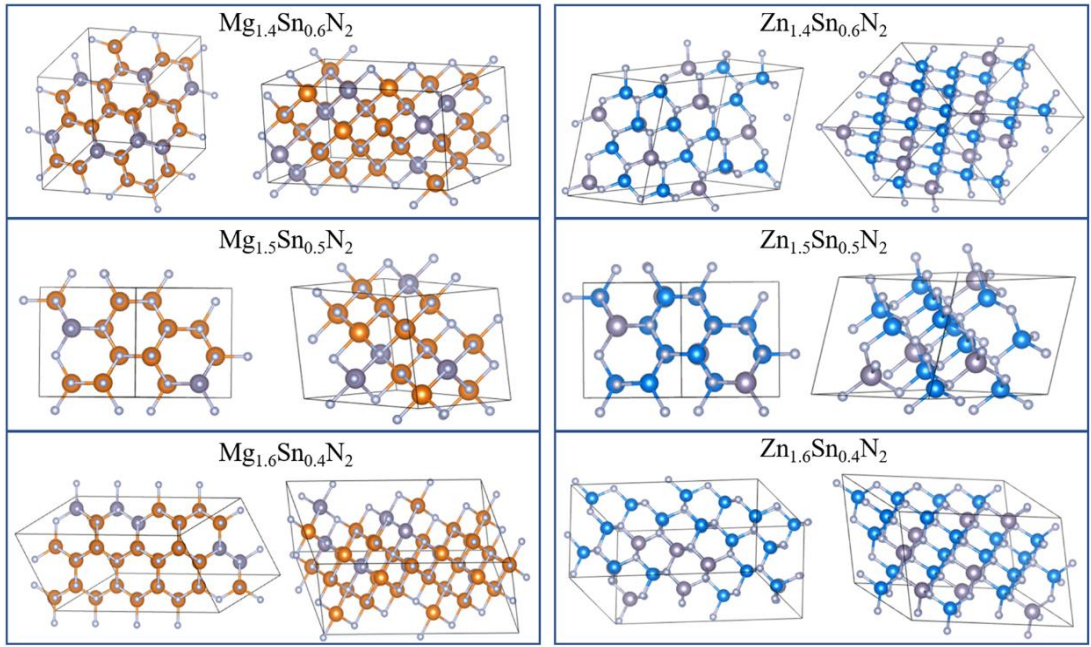

FIG. 7. Crystal structures of $Mg_{1+x}Sn_{1-x}N_2$ (left column) and $Zn_{1+x}Sn_{1-x}N_2$ (right column) with x=0.4, 0.5 and 0.6.

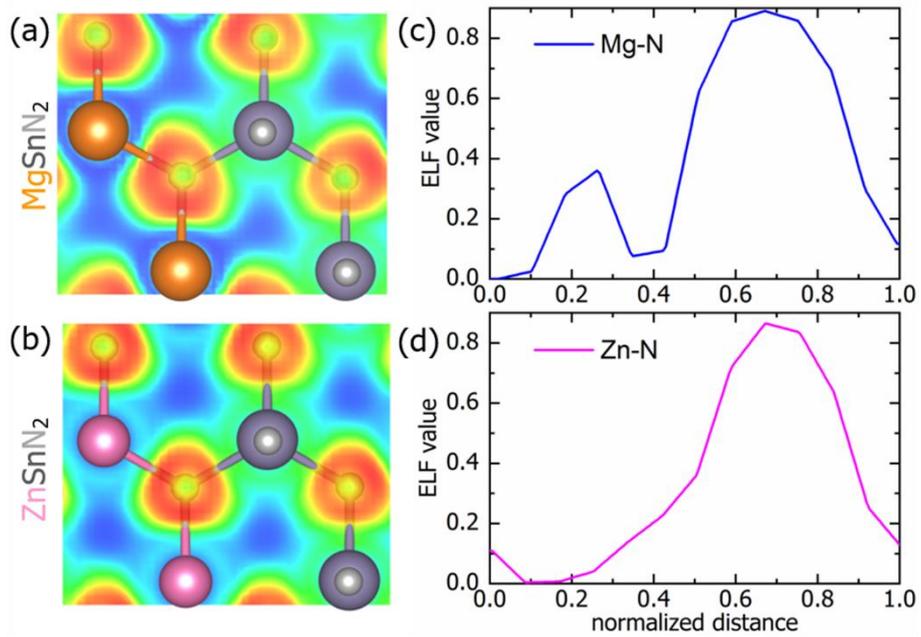

FIG. 8. 2D Electron localization function (ELF) contours and 1D profile of ELF along the Mg-N bond of $MgSnN_2$ (a-c) and Zn-N bond of $ZnSnN_2$ (b-d). Corresponding color code for constituents are shown on the left of parts a and b.

## B. Doping behavior

The off-stoichiometric Zn/Sn ratio influences the electrical properties of II-IV-nitrides [18, 23, 24, 26, 53]. As mentioned in Sec.I, ZnSnN$_2$ suffers from the degenerate electron concentration intrinsically, motivating research to lower its electron density for photovoltaic application [47]. It has been proposed that the compensation of Zn$_{Sn}$ and O$_N$ antisite defects form charge-neutral defect complexes, which would help lower the degenerate carrier concentration [54]. Moreover, it is experimentally demonstrated that both the Zn-rich condition and low-temperature growth are beneficial for suppressing the carrier concentration [23]. Notably, MgSnN$_2$ shows a similar behavior as ZnSnN$_2$, i.e., the electron density can reach $10^{20}$ cm$^{-3}$ [34]. Nonetheless, little is known theoretically about the doping behavior of Mg$_{1+x}$Sn$_{1-x}$N$_2$ to date.

Figure 9 displays the total density of states two ordered MgSnN$_2$ and Mg$_{1+x}$Sn$_{1-x}$N$_2$ with x = −0.4, −0.2, 0.2, 0.4, 0.5 and 0.6. As can been observed, Sn-rich compounds (Mg$_{0.6}$Sn$_{1.4}$N$_2$ and Mg$_{0.8}$Sn$_{1.2}$N$_2$) show n-type doping while Mg-rich samples (Mg$_{1.2}$Sn$_{0.8}$N$_2$, Mg$_{1.4}$Sn$_{0.6}$N$_2$, Mg$_{1.5}$Sn$_{0.5}$N$_2$ and Mg$_{1.6}$Sn$_{0.4}$N$_2$) exhibit p-type doping. Such a different doping behavior is induced by the antisite defects, namely, the formation of donor-type defect Sn$_{Mg}$ in Sn-rich samples and the acceptor-type defect Mg$_{Sn}$ in Mg-rich samples, which indicates that the electron transport properties can be tuned by varying the Mg/Sn ratio. Additionally, the formation of acceptor-type antisite defect Mg$_{Sn}$ can generate hole carriers once ionized, thus promoting the decrease of high electron carrier density in MgSnN$_2$, which agrees with the experimental observation that the electron density was reduced in Mg-rich layers [34].

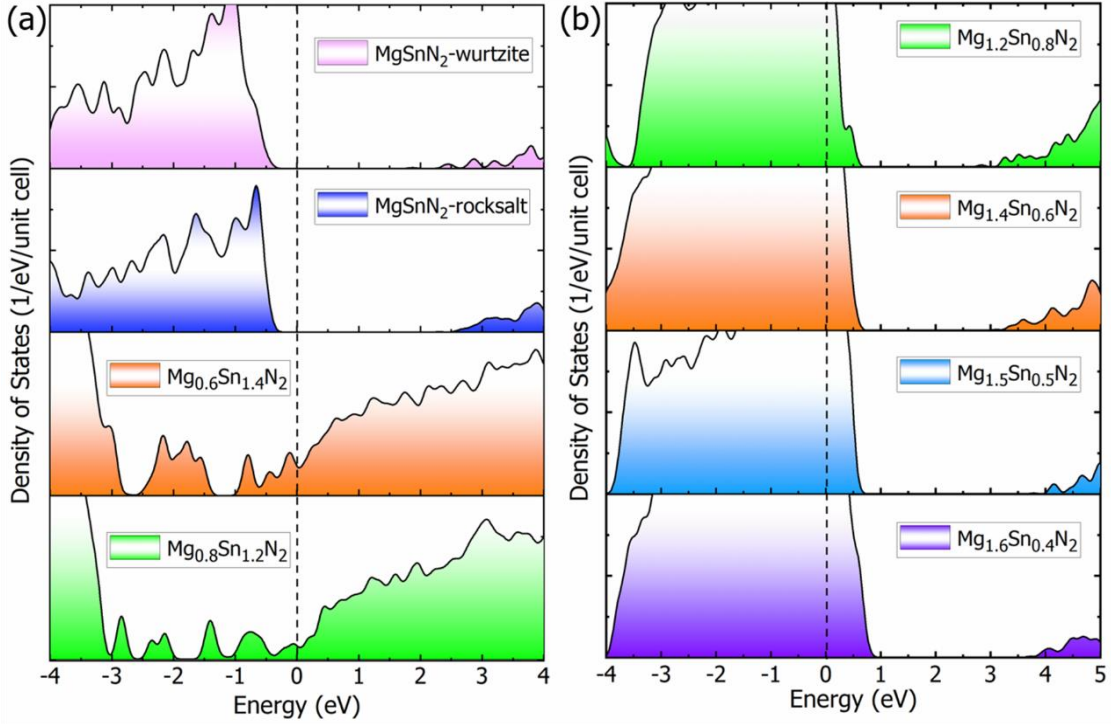

FIG. 9. Total density of states of (a) two ordered stoichiometric MgSnN$_2$, two crystallized Sn-rich compounds (Mg$_{0.6}$Sn$_{1.4}$N$_2$ and Mg$_{0.8}$Sn$_{1.2}$N$_2$), and (b) four crystallized Mg-rich compounds (Mg$_{1.2}$Sn$_{0.8}$N$_2$, Mg$_{1.4}$Sn$_{0.6}$N$_2$, Mg$_{1.5}$Sn$_{0.5}$N$_2$ and Mg$_{1.6}$Sn$_{0.4}$N$_2$). Fermi level is set to zero eV.

## V. CONCLUSIONS

We systematically investigated the impact of cation disorder in stoichiometric MgSnN$_2$ as well as the crystal structure and doping behavior in off-stoichiometric Mg$_{1+x}$Sn$_{1-x}$N$_2$ by means of the cluster expansion and first-principles calculations. Stoichiometric cation disorder in MgSnN$_2$ induces a decrease of the band gap because of a violation of the octet rule. We find that under nonequilibrium growth conditions, partially disordered MgSnN$_2$ cannot be avoided but fully disordered MgSnN$_2$ is difficult to be obtained. Moreover, the local disorder, namely, the formation of (4,0) or (0,4) tetrahedra, would lead to a large band gap reduction and impede the enhancement of the optical absorption. For off-stoichiometric Mg$_{1+x}$Sn$_{1-x}$N$_2$, in contrast to Zn$_{1+x}$Sn$_{1-x}$N$_2$, Mg-rich Mg$_{1+x}$Sn$_{1-x}$N$_2$ samples with x=0.4, 0.5 and 0.6 show penta-coordination, which can be ascribed to the higher ionicity of Mg-N bonds than Zn-N bonds as revealed by ELF. In addition, off-stoichiometric Mg$_{1+x}$Sn$_{1-x}$N$_2$ exhibits an ambipolar self-doping behavior, that is, Mg-rich samples show p-type doping while Sn-rich samples exhibit n-type doping. Mg-rich condition is suggested for the growth of MgSnN$_2$ since it would be helpful to lower the carrier concentration of MgSnN$_2$ for photovoltaic applications. Our work gives an understanding for basic physical properties stoichiometric and off-stoichiometric MgSnN$_2$ from the theoretical aspect, which should be useful for the further studies of MgSnN$_2$ based functional materials.

*Author to whom to correspondence should be directed:
Dan.Han@cup.uni-muenchen.de

ACKNOWLEDGMENTS

This work was funded by the Deutsche Forschungsgemeinschaft (DFG, German Research Foundation) under Germany's Excellence Strategy--EXC 2089/1--390776260.